\documentclass[twocolumn,eqsecnum,preprintnumbers,nofootinbib]{revtex4}
\usepackage{amsmath,amssymb,amsfonts,times,graphicx,color}
\pdfoutput=1

\newcommand{\beq}{\begin{equation}}
\newcommand{\eeq}{\end{equation}}
\def\bea{\begin{eqnarray}}
\def\eea{\end{eqnarray}}

\begin{document}

\title{A simple model of many-body localization}
\author{Brian Swingle}
\affiliation{Department of Physics, Harvard University, Cambridge MA 02138}

\date{\today}
\begin{abstract}
We study a simple and tractable model of many-body localization.  The main idea is to take a renormalization group perspective in which local entanglement is removed to reach a product state.  The model is built from a random local unitary which implements a real space renormalization procedure and a fixed point Hamiltonian with random exponentially decaying interactions.  We prove that every energy eigenstate is localized, that energy is not transported, and argue that despite being fine tuned, the model is stable to perturbations.  We also show that every energy eigenstate obeys an area law for entanglement entropy and we consider the dynamics of entanglement entropy under perturbations.  In the case of extensive pertubations we recover a logarithmic growth of entanglement observed in recent numerical simulations.
\end{abstract}

\maketitle

\section{Introduction}

Quantum systems can display the phenomenon of localization \cite{anderson_loc} where energy and charge are not transported under the dynamics of the system.  Localization was first understood in the context of single particle physics with static disorder \cite{anderson_loc}, but in many cases the inclusion of interactions or the coupling to a heat bath interrupts the localization physics and restores transport \cite{RevModPhys.57.287,PhysRevLett.95.206603,basko_loc}.  However, isolated local quantum systems may also enter a many-body localized phase where localization persists even with interactions \cite{basko_loc}.  Indeed, a many-body localized phase can exist when a local quantum system fails to thermalize and hence cannot serve as its own heat bath.  Note also that there can in principle be many kinds of many-body localized phases, but here we focus exclusively on a many-body localized phase in which all energy eigenstates are localized \cite{2007PhRvB..75o5111O}.  Much interest in this topic has been sparked by a beautiful series of papers including Refs. \cite{basko_loc,2007PhRvB..75o5111O,2009PhRvB..80k5104O,PhysRevB.82.174411}.  Our particular interest is in understanding the interplay between entanglement and renormalization in a many-body localized phase.

The original work of Anderson suggested that localization could persist in the presence of interactions \cite{anderson_loc}.  Basko et al. gave a picture of localization in Fock space as the physical basis for many-body localization \cite{basko_loc}.  Ref. \cite{2007PhRvB..75o5111O} suggested that in systems with a finite Hilbert space, all energy eigenstates could be localized in a many-body localized phase.  The phenomenology of entanglement dynamics in many-body localized phases has been explored in Refs. \cite{2008PhRvB..77f4426Z,2012PhRvL.109a7202B}.  Many-body localization in the context of driven systems was investigated in Refs. \cite{2013PhRvB..87m4202I,2013AnPhy.333...19D}.  Other interesting investigations include Refs. \cite{2012NJPh...14i5020C,2013EL....10137003D,2013NJPh...15d5021D,2013arXiv1304.1158H}.  Independent of the present work \footnote{This model was developed in part from conversations with M.P.A. Fisher and A. Chandran at ``Entanglement and Emergence II" at the Perimeter Institute. We were not aware at that time of the important related works of Refs. \cite{2013arXiv1304.4605S,2013arXiv1305.4915H,2013arXiv1305.5554S}.}, the dynamics of entanglement have been addressed in Refs. \cite{2013arXiv1304.4605S,2013arXiv1305.4915H,2013arXiv1305.5554S}.  Although our model was developed independently of Refs. \cite{2013arXiv1304.4605S,2013arXiv1305.4915H,2013arXiv1305.5554S}, it is easily seen that our fixed point Hamiltonian incorporates the dephasing dynamics first introduced in Ref. \cite{2013arXiv1304.4605S,2013arXiv1305.4915H}.  Finally, as this work was being finished, we learned of Ref. \cite{2013arXiv1306.5753B} where the idea of a local unitary is also used to elucidate the physics of localization.  Ref. \cite{2013arXiv1306.5753B} also contains an extensive numerical analysis, so our model complements their analysis.  To summarize, the physics we wish to describe is the existence of localized excited states, an entanglement area law for excited states, and the slow growth of entanglement under perturbations.

To this end we present a simple model of a many-body localized phase that can be straightforwardly analyzed.  We show that our model has the following features.  All many-body energy eigenstates are localized and obey an area law for entanglement entropy (defined below).  Entanglement does not spread under a localized perturbation up to exponentially small corrections.  Entanglement entropy displays a slow logarithmic growth for large regions at long times in the case of an extensive perturbation.  We also argue that the many-body energy spectrum has the expected exponentially small level spacing, that energy is not transported, and that the system is stable to small local perturbations.  Part of this stability argument is similar to that of Ref. \cite{basko_loc}.  We focus on entanglement because of the key role it has played in numerical studies and because it suggests a fruitful way to simulate many-body localized systems (using tensor networks \cite{mera,peps,terg} and DMRG \cite{PhysRevLett.69.2863,dmrg_review}) \cite{2008PhRvB..77f4426Z,2012PhRvL.109a7202B}.  Of course, in real experimental systems, various correlation functions will be much more accessible, but these can also be straightforwardly analyzed in our model.

Our perspective is strongly informed by renormalization group (RG) ideas and especially by the idea of real space renormalization using local unitaries (see for example Ref. \cite{vidal_er}).  A unitary $u$ is called local if it has the form $u = e^{-i k}$ for a local hermitian operator $k$ of bounded norm (not growing with system size).  Local unitaries are of interest because they can be simply approximated by discrete circuits of few-spin unitary operators using the Trotter product formula.  The key feature of such circuits in the case of local unitaries is that they have finite depth, that is they consist of only finitely many (not growing with system size) layers.  Hence such circuits preserve the locality properties of states and operators on which they act.

Much work has focused on the idea that certain such bounded depth circuits can be interpreted as effecting a real space coarse-graining transformation.  For example, in the context of ``entanglement renormalization" \cite{vidal_er}, one imagines using local unitaries to successively remove short-range entanglement from a state.  After each such step the degrees of freedom may be safely coarse-grained and the process can be repeated.  In this way, the entanglement in a quantum ground state, say, is displayed as a function of length scale.  The physical picture which emerges is that we remove local entanglement and coarse-grain until we reach a product state, either because the system is short-range entangled or because the system has a finite size.  There is growing evidence that this physical picture is a good one for ground states of local Hamiltonians.  This evidence comes from numerical simulations of various quantum critical points in one and two dimensions \cite{2009PhRvB..80k3103M,2009PhRvA..79d0301P}, from the realization that certain kinds of ground states have exact representations in this language \cite{PhysRevB.79.085118,PhysRevB.79.085119}, and more exotically, from the proposal \cite{PhysRevD.86.065007,2012arXiv1209.3304S} that such a picture of entanglement renormalization also partially underlies holographic dualities \cite{maldacena,polyakov,witten} in the context of quantum gravity.  In other words, a wide variety of ground states, including mean-field states, topologically ordered states, quantum critical points, and even states of gauge theories with holographic duals, can all be described in this unified language.  It is furthermore satisfying that this physical picture is also associated with numerical tools, namely tensor networks, that can be used to simulate such ground states \cite{dmrg_review}.

The picture just advocated works quite well so far for ground states, but the structure of entanglement in excited states is typically much more complex.  However, a many-body localized phase is crucially different.  There is a sense in which all energy eigenstates ``look the same" (for example, all eigenstates might be localized \cite{2007PhRvB..75o5111O}), so the story of entanglement renormalization in ground states should also apply to excited states in a many-body localized phase.  This observation forms the basis for our model.

Following the above RG perspective, a key component of our model is a local unitary $U$ which transforms exponentially localized states into exactly localized product states.  The other key component is a fixed point Hamiltonian $H_0$ which is diagonal in the local product basis and contains exponentially decaying interactions.  Our model should thus be interpreted as a renormalization group fixed point of a many-body localized phase, although we do not explicitly specify a renormalization group procedure beyond the local unitary $U$.  We note in passing that Refs. \cite{2010PhRvB..81m4202M,2013PhRvL.110f7204V} have explored possibly related RG perspectives.

We should also immediately mention that the model we introduce is not without imperfections.  For example, the Hamiltonian we analyze is not strictly local due to exponentially decaying interactions.  Also, we do not generically expect that all energy eigenstates are localized or that a single local unitary can reduce every energy eigenstate to an exact product state.  Thus our toy model is indeed fine tuned in the heuristic sense that all ``irrelevant" perturbations are zero\footnote{To be precise, we only know that an infinite class of perturbations have been set to zero.  The statement in the RG language that these perturbations are all irrelevant is the statement that the model is stable.}, but it does capture many of the phenomena expected in a many-body localized phase.

The rest of the paper is organized as follows.  First, we describe the basic model and establish that it is in a many-body localized phase in that energy is not transported.  Second, we analyze the physics of entanglement in the state.  Finally, we discuss the stability of the model and give a concluding discussion.

\section{Model}

Consider $N=L^d$ sites arranged on a hyper-cubic lattice in $d$ dimensions.  Each site $r$ carries a spin-half and we label the Pauli operators $X_r, Y_r, Z_r$.  We will use a renormalization group language to setup the model.  First, we define a ``fixed point" Hamiltonian in terms of projectors $P_r =(1-Z_r)/2$ as
\bea
&& H_0 = \sum_r J^1_r P_r + \sum_{rr'} J^2_{rr'} P_r P_{r'} \cr \nonumber \\
&& + \sum_{r r' r''} J^3_{r r'r''}P_{r}P_{r'}P_{r''} + ...
\eea
The couplings $J^i_{r_1 ... r_i}$ are drawn from a probability distribution $P[J;i,r_1 ... r_i]$. The couplings are assumed to decay exponentially with the distances $|r_a - r_b |$ and with increasing number of spins involved.  The length scale for the spatial decay is $\xi$, the localization length.  Throughout we denote by $J_0$ a typical short-ranged energy scale in $H_0$.

The Hamiltonian $H_0$ describes a perfectly localized state by construction.  Physically, we view it as the fixed point Hamiltonian obtained from a many-body localized system after some coarse-graining.  Clearly the probability distributions $P$ must be ``sufficiently random" for $H$ to truly describe a many-body localized state, otherwise adding small perturbations to $H$ could result in transport of energy under the perturbed dynamics, e.g. $P[J;1,r] = \delta(J-J_0)$ is clearly not suitable to describe a many-body localized state.  We further discuss these stability issues below.

We next model the coarse-graining process.  Introduce another local Hamiltonian $K$ which has strictly local interactions of bounded strength but is otherwise completely random.  Define the unitary operator
\beq
U = e^{- i K}
\eeq
which represents a fictitious time evolution for unit time under the Hamiltonian $K$.  We emphasize that this $K$ is not the physical Hamiltonian.  In fact, we wish to interpret $U$ is a renormalization group transformation which maps the physical exponentially localized many-body states to the ideal perfectly localized product states entering $H_0$.  To be precise, we define the full Hamiltonian
\beq
H = U H_0 U^{-1}
\eeq
where $H_0$ provides random energies and $U$ provides random localized wavefunctions.  We emphasize that the spectra of $H$ and $H_0$ are identical.  Also, although we do not consider it further here, we could include conserved charges by demanding that $K$ and $H_0$ commute with the conserved charge.

To show the locality of $H$, we use the machinery of Lieb-Robinson bounds \cite{LRbound}.  By assumption, $H_0$ consists of exponentially localized interactions.  Thus we need only show that the effects of $U$ are also local up to exponential corrections.  Since $K$ is a strictly local Hamiltonian built from Pauli operators with bounded coefficients, the Lieb-Robinson bound states \cite{LRbound} that there are constants $\xi_{LR}$ (a length), $v_{LR}$ (a velocity), and $C$ such that for two commuting operators $A$ and $B$ separated by a distance $d$ we have
\beq
|| [A(t),B(0)] || <  C ||A || || B|| e^{-(d-v_{LR} t)/\xi_{LR}}.
\eeq 
where $A(t) = e^{i K t} A e^{- i K t}$ and where $||O||$ is the operator norm of $O$.  This bound implies that for times of order one (e.g., for transformations by $U=e^{-iK}$), distant operators still commute up to exponential corrections.  The length $\xi_{LR}$ plays the role of a localization length, and we henceforth restrict to $K$ such that the Lieb-Robinson bound is tight with $\xi_{LR} = \xi$.  Thus the Hamiltonian $H$ consists of exponentially decaying interactions with characteristic length scale $\xi$ as desired.  It should be noted that from one point of view our model is unrealistic in that it is not strictly local, that is it contains exponentially decaying (and not strictly local) interactions.

Let us also observe that the many-body spectrum of $H$, which is identical by construction to that of $H_0$, is generic in the sense that the level spacing is typically of order $2^{-N}$ where $N=L^d$ is the number of spins.  This follows from the average density of states.  For example, suppose $P[J,i>1]=0$ so that one has only random onsite energies with $P[J,1] = \theta(J_0 - J ) \theta(J) / J_0$.  In this case the many-body spectrum is $E(n) = \sum_r J_r n_r$ with $n_r = 0,1$.  The density of states is 
\beq
D(E) = \sum_n \delta(E-E(n)),
\eeq
and the disorder averaged density of states is
\beq
\overline{D(E)} = \sum_n \overline{\delta(E-E(n))}.
\eeq
To compute the average we simply use that the onsite energies are uncorrelated, so the central limit theorem gives, for $\tilde{N}(n) = \sum_r n_r \gg 1$, 
\beq
p(E(\tilde{N})) \propto \exp{\left(-\frac{(E-\tilde{N} J_0/2)^2}{2 \tilde{N} (J_0^2/4)} \right)}.
\eeq
The number of $n_r$ summing up to $\tilde{N}$ is $\left(\begin{array}{c}
                                                          N \\
                                                          \tilde{N} 
                                                        \end{array}
\right),$ and hence we have roughly
\beq
\overline{D(E)} \sim \int d\tilde{N} \left(\begin{array}{c}
                                                          N \\
                                                          \tilde{N} 
                                                        \end{array}
\right) \exp{\left(-\frac{(E-\tilde{N} J_0/2)^2}{2 \tilde{N} (J_0^2/4)} \right)}.
\eeq
The $\tilde{N}$ integral is strongly peaked provided $2 E/J_0\sim N$ in which case $\overline{D(E)} \sim 2^N$ as claimed.  Adding in further neighbor interactions does not substantially change the above reasoning.

The Hamiltonian $H$ describes a many-body localized system because energy is not transported.  There are two results which illustrate this point.  First, given any two many-body eigenstates that differ by spin flips in distant regions, the local physics of these states will be nearly identical away from the regions where the spin flips occurred.  Indeed, the operator $U X_r U^{-1}$ moves from one many-body energy eigenstate to another and is manifestly exponentially localized due to the localized structure of $U$.  This means that adding a precise amount of energy to a state is a localized operation.

Second, suppose we start with a many-body eigenstate $|E\rangle$ and perturb the system locally by acting with a local operator $O(r_0)$.  Then the amplitude to find the system in another many-body eigenstate associated with a distant eigenstate, e.g. $U X_r U^{-1} |E\rangle$, is exponentially small as $|r-r_0| \rightarrow \infty$.  This is true for all time since $U$ is localized and the dynamics of $H_0$ can only lead to dephasing and not distant spin flips.  Hence the amplitude of a distant spin flip can be bounded at all times by a constant which decays exponentially with distance.  Thus energy is not transported.  For example, in a process where we start in the ground state, inject energy at $r_0$, and try extract energy at $r_1$, we only succeed with exponentially small probability.

\section{Entanglement}

We now obtain results about the structure of entanglement in our model.  Entanglement entropy is defined by splitting the Hilbert space into two pieces associated with a region $A$ and its complement $B$.  The Hilbert space factorizes as $\mathcal{H} = \mathcal{H}_A \otimes \mathcal{H}_B$ and the state of region $A$ may be obtained from the state of the whole system via a partial trace over $B$:
\beq
\rho_A =\text{tr}_B(\rho_{AB}).
\eeq
The entanglement entropy $S(A)$ of $A$ is then the von Neumann entropy of $\rho_A$:
\beq
S(A) = - \text{tr}_A(\rho_A \ln{(\rho_A)}).
\eeq
When $\rho_{AB}$ is a pure state, $S(A)$ indeed measures entanglement between $A$ and $B$.

\subsection{Area law for energy eigenstates}

We first address entanglement entropy for energy eigenstates.  To begin, every eigenstate of $H_0$ is a product state and hence is unentangled.  Every eigenstate of $H$ is obtained from an eigenstate of $H_0$ by acting with $U$: $|E\rangle = U \prod_r |n_r \rangle$ where $P_r |n_r \rangle = n_r |n_r \rangle$.  Hence any entanglement present in $|E\rangle$ is due to the action of $U$.  $U$ is generated by a fictitious time evolution with Hamiltonian $K$ for a time $t=1$, and we may bound the entanglement generated by $U$ because $K$ is local.

Introduce the region $A$ whose entanglement we wish to compute and its complement $B$.  $K$ may be decomposed as
\beq
K = K_A + K_B + \sum_\alpha k_\alpha
\eeq
into terms acting only on $A$ or $B$ and $AB$ interaction terms of the form $k_\alpha = O^\alpha_A \otimes O^\alpha_B$ where $O_R$ acts only on region $R$.  $K_A$ and $K_B$ do not directly generate entanglement, so we must only consider the $AB$ interaction terms in $K$.  The strict locality of $K$ implies that the number of such terms grows only with the boundary size $|\partial A|$ of $A$.  The entangling power of each such term is rigorously bounded by
\beq
\frac{dS(A)}{dt} \leq c ||O^\alpha_A || ||O^\alpha_B||
\eeq
with $c$ a numerical constant \cite{ent_cap_HAHB,PhysRevLett.97.150404,PhysRevLett.97.050401}.  Intuitively this is because the maximum possible entropy that a unitary of the form $e^{-i k_\alpha t}$ can add to region $A$ is bounded (by the log of the dimension $d_\alpha$ of the Hilbert space on which it acts) with the maximum typically occuring when $t = t_{max} \sim ||k_\alpha||^{-1}$ (an inverse energy).  Thus heuristically one has
\beq
\frac{dS}{dt} \sim \frac{\Delta S_{max}}{t_{max}} \sim \ln{(d_\alpha)} ||k_\alpha|| = \ln{(d_\alpha)} ||O^\alpha_A || || O^\alpha_B ||.
\eeq

To treat the sum over $\alpha$ in $K$ we use the the Trotter product formula to show that
\beq
\frac{dS(A)}{dt} \leq \sum_\alpha c ||O^\alpha_A || ||O^\alpha_B||.
\eeq
Since the total number of such operators is proportional to $|\partial A|$ and since the all interactions have bounded strength (meaning $||k_\alpha || < \tilde{k} $ for some fixed number $\tilde{k}$), the total entropy generated by time evolution for time $t=1$ is bounded
\beq
\Delta S(A) \leq c' |\partial A|.
\eeq
Hence every many-body energy eigenstate obeys an area law for entanglement entropy.

\subsection{Time dependent states}

We can also consider time dependent states.  Suppose first that we make a local perturbation starting from an energy eigenstate $|E\rangle$.  As usual, we use $U$ to translate this perturbation into a state evolving under $H_0$.  If the perturbation happens to push the system into another energy eigenstate then nothing will happen.  If the perturbation produces a superposition of energy eigenstates then $H_0$ will rapidly decohere the superposition.  However, as with energy transport, the maximum amplitude to produce distant excitations is bounded by an exponentially decaying envelope.  If $\psi(n)$ are the amplitudes of the eigenstates $|n\rangle$ of $H_0$ then the maximum entropy of a region $A$ is the entropy of the density matrix
\beq
\rho_A(n_A, n'_A) = \sum_{n_B} \psi(n_A n_B)\psi^*(n'_A n_B)
\eeq
plus the maximum entropy $U$ can add.  But by assumption the amplitudes $\psi$ factorize except neat the localized perturbation and cannot give a contribution to the entropy which grows with region size.  Hence the entropy of any region $A$ sufficiently larger than the size of the local perturbation will be fixed at its value in the initial eigenstate up to exponentially decaying corrections.

We can also consider extensive perturbations.  Suppose we begin with a global product state.  We use $U$ to translate this into a state which evolves under $H_0$.  Because $U$ is local, the resulting initial state is short range correlated and has at most area law entanglement.  The initial state will be a superposition of different eigenstates of $H_0$.  Due to the exponentially decaying interactions in $H_0$, the initial state will slowly lose coherence and entanglement will be generated.  

We give a simple model calculation which illustrates the effect.  Consider two spins with Hamiltonian
\beq
h = J_1 P_1 + J_2 P_2 + J_{12} P_1 P_2
\eeq
and suppose for simplicity that the spins begin in identical product states
\beq
|\psi(0) \rangle = (\alpha |0\rangle + \beta |1\rangle )^{\otimes 2}.
\eeq
This state evolves into the state
\bea
&& |\psi(t)\rangle = e^{-i h t} |\psi(0)\rangle = \alpha^2 |00\rangle + \alpha \beta e^{-i J_2 t} |0 1 \rangle \cr \nonumber \\
&& + \beta \alpha e^{-i J_1 t} |1 0\rangle + \beta^2 e^{-i (J_1+J_2 + J_{12})t} |11 \rangle
\eea
after time $t$.  For concreteness, suppose we have initially that $\alpha=\beta=1/\sqrt{2}$.  Then the reduced density matrix of spin one takes the form
\beq
\rho_1(t) = \frac{1}{2} \left(\begin{array}{cc}
                  1 & \frac{e^{i J_1 t}\left(1+e^{i J_{12} t}\right)}{2} \\
                  \frac{e^{-i J_1 t}\left(1+e^{-i J_{12} t}\right)}{2} & 1 
                \end{array}
\right).
\eeq
Hence we see that, independent of $J_1$ and $J_2$, the state becomes maximally mixed when $J_{12} t = \pi$.  Of course, in this simple model the system returns to a pure state after the same time, but in the Hamiltonian $H_0$ there are additionally many other spins which also interact with a given spin.  These additional interactions lead to decoherence which implies that once purity is lost, it is not recovered (modulo recurrences on time scales of order $2^N$).

Returning to the main problem, consider again a region $A$ evolving from an initial product state for a time $t$.  The physical picture is that at time $t$ entanglement has been generated between $A$ and spins a distance
\beq
\ell(t) \sim \xi \ln{\left(J_0 t/\pi\right)}
\eeq
away.  This is because, assuming exponentially decaying interactions and following the model calculation above, only these spins have had enough time to entangle with region $A$.  Several types of behavior are then possible.  Of course, there can be some initial transient behavior before the localization scale is reached, but afterwards, if the characteristic linear size $L$ of $A$ is much greater than $\ell(t)$, then we expect an entropy going like $S(A,t) \sim |\partial A| \ell(t)$. 

Eventually the entropy will saturate since we have the bound
\beq
S(A) \leq N_A \ln{2}.
\eeq
In fact, the entropy can never exceed the entropy of the ``diagonal" state consisting of $\rho_A$ with all off diagonal elements in the $Z$ basis set to zero.  Indeed, this process of setting off diagonal elements to zero amounts to a measurement of $Z$ for every spin in $A$,
\bea
&& \rho_A \rightarrow \mathcal{E}(\rho_A) = \sum_{\{x_r=0,1\}} M(x) \rho_A M^\dagger(x),
\eea
where
\beq
M(x) = \left(\prod_r (P_r)^{x_r} (1-P_r)^{1-x_r}\right)
\eeq
so that the set $\{M(x)\}$ represents a projective measurement of all $Z$ variables.  Such a measurement always increases the entropy.  Hence the diagonal ensemble upper bounds the entanglement entropy.

\subsection{Justification of entanglement growth rate}

We now give a more rigorous jutification of the above results.  First, since $U$ can at most add an area law worth of entanglement to a state, for the purposes of asymptotic time dependence, we may as well work directly with a state evolving under $H_0$.  $H_0$ has two very special properties that facilitate the analysis.  First, the various terms in $H_0$ are all diagonal in a local product basis.  Second, every term in $H_0$ commutes with every other term.  This implies that the time evolution generated by $H_0$, call it $W(t) = e^{- i H_0 t}$, can be factorized as
\beq
W = W_A W_B W_{AB},
\eeq
where each term acts on the indicated subsystem and all terms commute.

If $\rho_{AB}(0)$ is the initial state of $AB$, then we wish to compute the time dependence of the state of $A$ defined as
\beq
\rho_A(t) = \text{tr}_B (W(t) \rho_{AB}(0) W^\dagger(t)).
\eeq
We immediately obtain 
\beq
\rho_A(t) = W_A \text{tr}(W_{AB}\rho_{AB}(0) W^\dagger_{AB}) W^\dagger_A
\eeq
where we have used the the cyclic property of the trace to remove $W_B$.  Furthermore, the unitary $W_A$ doesn't change the spectrum of $\rho_A$, so it cannot effect the entanglement entropy.  Hence we may as well set $W_A = W_B = 1$, so that the whole entropy is manifestly dependent only on the $AB$ interactions in $W_{AB}$ and the initial state.

Suppose now that, unlike in our model, the interactions between $A$ and $B$ were of strictly finite range.  Then we see that the $W_{AB}$ acts on a finite Hilbert space near the boundary of $A$ of dimension roughly $2^{R |\partial A|}$ where $R$ is the range of the terms in the interaction Hamiltonian $H_{AB}$.  Since the amount of entropy such a unitary can add to a system is bounded by the logarithm of the dimension of the Hilbert space on which it acts, we see that such a $W_{AB}$ with short-range interactions cannot add more than roughly $R |\partial A|$ entropy.  Hence a strict area law will be obeyed for sufficiently large regions at all times.  In fact, in one dimension the situation is even worse.  There, if the range $R$ is finite, then $W_{AB}(t)$ is quasiperiodic in $t$ and the entanglement entropy actually suffers recurrences on time scales of order $1/J_0$.

Thus we see that exponentially decaying interactions are essential if we want to have any sustained growth of entanglement for large regions.  The virtue of the above argument is that it immediately provides an estimate of the entropy growth.  We must ask what is the effective range, $R_{\text{eff}}$, of $H_{AB}$ at time $t$.  Since $W_{AB} = e^{-i H_{AB} t}$ and since the terms in $H_{AB}$ between spins a distance $r$ apart are roughly of order $J_0 e^{-r/\xi}$, the natural choice is to say that $R_{\text{eff}}(t)$ is such that
\beq
t J_0 e^{-R_{\text{eff}}/\xi} \sim 1.
\eeq

Now because $Z_r$ commutes with $H_0$, the diagonal elements of $\rho_A$ in the $Z$ basis do not change with time.  As we already discussed, the entropy of the state with all the off-diagonal elements zero bounds the entropy of $\rho_A$ itself.  But more than this, since we expect dephasing to occur, the late time state should indeed be close to this $Z$-measured state. Let $s$ be the entropy per spin in the dephased state (it can be much less than one and depends on the initial conditions).  Then the entropy of region $A$ should grow as
\beq
S(A,t) = s |\partial A| R_{\text{eff}}(t).
\eeq
Since $R_{\text{eff}}(t) = \xi \ln{(J_0 t)}$ we indeed recover the claimed logarithmic growth.  In fact, with the bounds on entanglement growth used above, we can prove that the entanglement cannot grow faster than this estimate.  Furthermore, because the Hamiltonian is generic we expect it to saturate this bound on the rate of entanglement growth.

\section{Stability}

The model we have presented is meant to represent a system deep within a many-body localized phase.  However, as we already mentioned, a system should only be called many-body localized if it is stable to small perturbations.  Thus in order for $H$ to describe a many-body localized system, it must be that $H' = H + g V$ has all the same universal physics as $H$ for all local perturbations $V$, e.g. area laws for excited states and slow growth of entanglement, provided $g$ is small enough.  If, for example, we choose the couplings in $H_0$ and the generator $K$ to be non-random, then we would certainly not be describing a many-body localized phases since perturbations would immediately lead to transport of energy.  In this case we might say that the ``fixed point" described by $H$ is infinitely unstable, i.e. has infinitely many relevant perturbations, even though $H$ is indeed localized by construction.

What follows in this section are heuristic arguments and some partial results towards a proof of stability.  We first describe the basic intuition and then discuss a more technical approach based on adiabatic continuity.

Assuming that $H_0$ and $K$ are random as described above, we wish to analyze the effects of perturbations $g V$ with $V$ local and $g$ small.  The perturbed Hamiltonian is
\beq
H' = H + g V,
\eeq
or after transforming by the local unitary $U$,
\beq
H_0' = H_0 + g U^{-1} V U = H_0 + g V_0.
\eeq
The crucial point is that $V_0$ is still local up to exponentially decaying corrections.  At this point, we certainly cannot prove that the many-body localized state is stable, and indeed, there are many possible subtleties, for example, a few many-body eigenstates might delocalize while the whole system effectively remains in a many-body localized phase.  However, a some perturbative intuition that suggests the model is stable.

Indeed, since $V_0$ is a local operator, it will typically only connect states that differ by a finite energy (not decreasing with system size).  If we assume that the terms in $V_0$ have bounded norm, then the matrix elements are of order $g$ or smaller and decay exponentially with distance.  If $V_0 = \sum_r v_r$ with each $v_r$ a local operator and if $|n\rangle$ and $|m\rangle $ are two eigenstates of $H_0$, then matrix elements of the form $\langle n |v_r |m\rangle$ will vanish exponentially unless $n$ and $m$ differ only by changes near site $r$.  However, in this case $E(n)$ will typically differ from $E(m)$ by a finite energy, say of order $J_0$, and hence the perturbing matrix elements will be quite small compared to the diagonal elements provided $g$ is small.

Of course, we can use $V_0$ to connect states which are closer in energy by going to higher orders in perturbation theory, but these processes will be exponentially suppressed due to the large number of off resonant intermediate states required.  The picture which thus emerges is that we must go to an exponentially suppressed high order in perturbation theory or make use of an exponentially small interaction to connect states with nearly degenerate energies.  This scenario is then essentially the picture advocated by Ref. \cite{basko_loc} where one has localization in the many-body Fock space.  It was argued in Ref. \cite{basko_loc} that the growth in the number of terms at each order of perturbation theory is commensurate with an interpretation as an Anderson model in Fock space with bounded connectivity.  More concretely, if free particle energy eigenstates are points and if two points share an edge provided they are connected by a matrix element of the interaction, then the resulting graph has an Anderson transition if the connectivity is finite.  It would be interesting to carry out a more quantitative analysis of this scenario in our model, but in this work we turn to different but related point of view.  

We will attack the problem from the point of view of adiabatic continuity.  Ref. \cite{Altshuler13072010} has analyzed the quantum adiabatic algorithm for a different model and concluded that Anderson localization makes the adiabatic algorithm fail (due to an extremely small gap).  However, our perspective will be somewhat different in that we will not require that ground states be mapped to ground states but only that energy eigenstates be mapped to energy eigenstates.

\subsection{Adiabatic continuity}

Let us begin by recalling that for band insulators (not disordered), the ground state is localized and hence the linear response conductivity vanishes at zero temperature.  Of course, at any finite temperature the system will conduct due to a finite density of delocalized quasiparticles.  However, the important point is that the localization of the ground state persists as the Hamiltonian is smoothly changed.  Hence the conductivity vanishes at zero temperature everywhere within the phase.  On the other hand, highly excited energy eigenstates are not localized and an applied field leads to a non-zero response.  From the point of view of adiabatic continuity, the ground state of, say, a band insulator, differs dramatically from a highly excited state in that the ground state is separated by a finite gap from other eigenstates while a highly excited state typically sits in a nearly continuous set of states with a level spacing going like $e^{-N}$.  Furthermore, these nearly degenerate states can be easily connected to each other by local operators.

In the context of ground states of band insulators, the basic adiabatic argument runs as follows.  Consider a family of Hamiltonians $H(g)$ each possessing a localized ground state $|\psi(g)\rangle$ and a gap $\Delta(g) \geq \tilde{\Delta}$.  Suppose we start at the initial ground state $|\psi(0)\rangle $ and time evolve under the Hamiltonian $H(g(t))$ for some slowly varying function $g(t)$:
\beq
\frac{d |\phi(t)\rangle}{dt} = H(g(t)) |\phi(t)\rangle
\eeq
and
\beq
|\phi(t=0)\rangle = |\psi(g=0)\rangle.
\eeq
Then because $H(g)$ is gapped for all $g$ with minimum gap $\tilde{\Delta}$, it follows that local properties of $|\psi(1)\rangle$ will be accurately reproduced provided $\frac{1}{g} \frac{dg}{dt} \ll \tilde{\Delta}$.  In general, to truly make the time evolved state $|\phi(t)\rangle$ close to $|\psi(1)\rangle$ in the sense of having high overlap, we must evolve for time growing with system size, but if we really want all global properties to be preserved then we can use a modified procedure known as quasi-adiabatic continuation \cite{quasiadiabatic} discussed below.  Regardless, as far local properties are concerned the adiabatic argument works well.  However, we see immediately that a key difference for excited states is the complete lack of a gap between nearby energy eigenstates.  Since these eigenstates can be connected to each other via local operators, it follows that time evolution under a local Hamiltonian for any reasonable amount of time, e.g., less than times of order $e^N$, will strongly mix various excited states.  Hence we apparently cannot learn as much about excited states using adiabatic continuity.

The situation is quite different in a many-body localized phase.  We have already shown that while the level spacing of the many-body spectrum goes like $e^{-N}$ near the center of the spectrum, states with nearly degenerate energies are typically associated with very different configurations of spins.  As argued in the perturbative discussion above, local operators typically only connect states with very different energies.  Hence as far as local operators are concerned, energy eigenstates in a many-body localized state typically have an effective gap to other eigenstates.  In other words, states $|E \rangle $ and $|E'\rangle$ with a sizable matrix element $\langle E | O |E'\rangle$ for some localized operator $O$ typically differ in energy by a relatively large amount, say, typically $|E-E'| \geq \Delta(O)$ where the effective gap $\Delta(O)$ depends on how localized the operator $O$ is.

We now repeat the dynamical evolution
\beq
\frac{d |\phi(t)\rangle}{dt} = H(g(t)) |\phi(t)\rangle,
\eeq
with
\beq
|\phi(t=0)\rangle = |E\rangle
\eeq
where $|E\rangle $ is an arbitrary many-body localized energy eigenstate.  Then as we argued above, since $H(g(t))$ is local for all $t$ and since there is an effective gap for local perturbations, the adiabatic evolution maps eigenstates to eigenstates in a many-body localized phase.  This should be compared with the result just for ground states in the case of conventional insulators.

Then using the Lieb-Robinson arguments above, it follows that the unitary $Q(t_{ad})$ which effects the adiabatic time evolution for time $t_{ad}$ is a local unitary up to exponentially decaying corrections.  Hence the new energy eigenstates may be approximated by $Q(t_{ad}) |E\rangle $, but since $|E\rangle$ is itself related to a product state by $U$, we see that the new energy eigenstates are also related to product states by the local unitary $Q(t_{ad}) U$ and thus are localized.  Hence the fact that energy eigenstates of the final Hamiltonian $H(1)$ are localized can be heuristically derived from the fact that the energy eigenstates of the initial Hamiltonian $H(0)$ are localized provided the effective gap $\Delta(O)$ for local perturbations doesn't collapse.

\subsection{Quasi-adiabatic continuity}

Having given various intuitive arguments for stability, we now want to formalize as much as possible the problem of proving stability.  The main tool we use in the remainder of this section is Hastings' quasi-adiabatic continuation \cite{quasiadiabatic}.  Our considerations are also potentially related to Hastings' recent work on adiabatic continuity in disordered systems \cite{2010arXiv1001.5280H}.

To achieve high overlap between an adiabatically evolved state and the target, one must evolve the system for a time which grows with system size.  The following simple example illustrates this need. Suppose we have a single spin with Hamiltonian $h(t)$ which we evolve for time $t_{ad}$.  Let $|\phi(t)\rangle_1 $ denote the result of adiabatic time evolution and let $|\psi(t)\rangle_1$ denote the instantaneous ground state of $h(t)$.  Suppose that $h(t)$ changes slowly enough so that
\beq
\langle \phi(t)|\psi(t)\rangle_1 = 1-\epsilon.
\eeq
If we now perform the same adiabatic evolution in a many-body system of $N$ such spins, all with the same hamiltonian and not interacting, then we have
\beq
|\phi(t)\rangle_N = |\phi(t)\rangle_1^{\otimes N}
\eeq
and
\beq
|\psi(t)\rangle_N = |\psi(t)\rangle_1^{\otimes N}.
\eeq
Thus the overlap is
\beq
\langle \phi(t)|\psi(t)\rangle_N = (1-\epsilon)^N
\eeq
because we have an independent probability of error at each site.  Hence to make $(1-\epsilon)^N$ close to one for $N$ large, $\epsilon$ must be chosen to decrease with $N$.

Remarkably, it is possible to construct an exponentially localized hermitian operator which generates a fictitious time evolution that exactly reproduces the ground state in a gapped system.  This method is called quasi-adiabatic evolution and was pioneered by Hastings \cite{quasiadiabatic}.  Consider a Hamiltonian $H(s)$ which depends on a parameter $s$.  We would like to find a local operator $J(s)$ such that the solution of
\beq
i \partial_s |\phi(s)\rangle = J(s) |\phi(s)\rangle
\eeq
has $|\phi(s)\rangle$ equal to the exact ground state of $H(s)$ for all $s \in [0,1]$.

The ground state of $H(s)$ is denoted $|\psi_0(s)\rangle$.  We assume the system has gap $\Delta(s) = E_1(s) - E_0(s)$ which is bounded from below by an $s$ and system size independent constant $\Delta(s) > \Delta$.  Let $F(t)$ be a fast decaying function of $t$ with the following properties.  First, its Fourier transform $\tilde{F}$ satisfies $\tilde{F}(\omega) = -\frac{1}{\omega}$ for $|\omega| \geq \Delta $ and second, $\tilde{F}(\omega=0)=0$.  Now we define the generator $J(s)$ of quasi-adiabatic evolution as
\beq
-i J(s) = \int_{-\infty}^\infty dt F(t) e^{i H(s) t} \partial_s H(s) e^{-i H(s) t}.
\eeq
If we now apply $J(s)$ to the ground state we find
\bea
&& -i J(s) |\psi_0(s)\rangle = \sum_n \tilde{F}(E_n-E_0) | \psi_n(s) \rangle \cr \nonumber \\
&& \times \langle \psi_n(s) | \partial_s H(s) |\psi_0(s) \rangle,
\eea
but this last expression, using the properties of $\tilde{F}$, is simply
\bea
&& \sum_{n\neq0} \frac{-1}{E_n-E_0} | \psi_n(s) \rangle \langle \psi_n(s) | \partial_s H(s) |\psi_0(s) \rangle\cr \nonumber \\
&& = \partial_s |\psi_0(s)\rangle.
\eea
The last equality above is conventional perturbation theory.  Thus we have that
\beq
i \partial_s |\psi_0(s) \rangle = J(s) |\psi_0(s)\rangle
\eeq
with $J(s)$ a quasi-local hermitian effective Hamiltonian which generates the quasi-adiabatic evolution.

The quasi-locality of $J(s)$ follows from the Lieb-Robinson bound and from the fast decay of $F(t)$ with $t$.  In brief, $e^{i H(s) t} O e^{-i H(s) t}$ is localized up to exponential corrections beyond the Lieb-Robinson light cone $v_{LR} t$, and although we integrate over all $t$, the integral is weighted with a fast decaying function $F$.  Large values of $t$ are almost exponentially suppressed (faster than any power) and since small values of $t$ lead to exponentially localized operators, we see that every term in $\partial_s H_s$, even after the integral transform is applied, remains local up lengths of roughly $v_{LR}/\Delta$ with nearly exponentially decaying corrections beyond.  To formalize these statements, let $\partial_s H(s) = \sum_r V_r(s)$, i.e. a sum over local terms $V_r(s)$.  Now consider $J_r(s,T)$ defined by
\beq
-i J_r(s,T) = \int^T_{-T} dt F(t) e^{i H(s) t} V_r(s) e^{-i H(s) t}.
\eeq
The Lieb-Robinson bound states that $J_r(s,T)$ is localized to within $v_{LR} T$ of $r$ up to an exponentially small error, but we also know from the fast decay of $F$ that
\beq
|| J_r(s,T) - J_r(s,\infty) || < F_\infty(\Delta T) ||V_r(s)||
\eeq
with $F_\infty(x)$ another fast decaying function related to $F$.  For example, we could expect to have $F_\infty(T \Delta) \sim e^{-T\Delta}$ or nearly exponential decay.  Thus $J_r(s,T)$ is localized up to exponential error, $J_r(s,T)$ is exponentially close to $J_r(s,\infty)$, and $\sum_r J_r(s,\infty)$ is nothing but our original $J(s)$.

It is interesting to instructive to apply the quasi-adiabatic technology to the non-interacting many spin problem from the beginning of Appendix A.  In this case we see immediately that $J(s)$ is strictly local and in fact is a sum of independent terms for each spin.  Each such term is identical and by construction perfectly evolves ground states into ground states.  Indeed, in this case it also perfectly evolves excited states into excited states since each spin has only two levels and unitary evolution preserves orthogonality.

\subsection{Qausi-adiabatic evolution for many-body localization}

Since the evolution generated by $J(s)$ is quasi-local, it follows that if one ground state is localized then so is every other ground state.  Thus it would be very useful if we could apply quasi-adiabatic continuation to more general localized states.  We can try to repeat the same story for a many-body localized phase, but the immediate problem is that the many-body localized Hamiltonian does not possess a gap (around the ground state or around excited states).  However, as discussed above, there is a sense in which each energy eigenstate has a gap to other eigenstates which can be connected to it by a local operator.  In this sense, all eigenstates of a many-body localized phase should be treated on the same footing.  For example, in our model all eigenstates manifestly share the same properties and each could be the ground state of a local Hamiltonian.

Considering again a parameter dependent Hamiltonian $H(s)$ with a spectrum in which every energy level is distinct and separated by a gap of at least $\gamma$ from every other level.  Let the normalized energy eigenstates be $|\psi_n(s)\rangle$.  By definition we have
\beq
H(s) |\psi_n(s)\rangle = E_n(s) |\psi_n(s)\rangle.
\eeq
Upon differentiating both sides with respect to $s$ we find (denoting $\partial_s$ with a $'$)
\beq
H' |\psi_n \rangle + H |\psi_n\rangle' = E_n' |\psi\rangle + E_n |\psi_n\rangle' .
\eeq
Since $|\psi_n(s)\rangle $ is normalized for all $s$, we have $\langle \psi_n | (|\psi_n \rangle') = 0$. Thus we can simplify the equation by projecting onto $|\psi_n\rangle $ and the complement.  Projecting onto $|\psi_n\rangle$ gives
\beq
E_n' = \langle \psi_n | H' | \psi_n \rangle.
\eeq
Projecting onto the complement gives
\beq
(H - E_n) |\psi_n\rangle' = - (1- |\psi_n \rangle \langle \psi_n |) H' |\psi_n\rangle.
\eeq
Since $H-E_n$ is invertible on the the complement to $|\psi_n\rangle$ we have an equation for $|\psi_n\rangle '$
\beq
|\psi_n\rangle' = - (H-E_n)^{-1}(1- |\psi_n \rangle \langle \psi_n |) H' |\psi_n\rangle.
\eeq

We now apply $J(s)$ as defined above to a general energy eigenstate to obtain
\bea
&& -i J(s) |\psi_n(s)\rangle = \sum_m \tilde{F}(E_m-E_n) | \psi_m(s) \rangle \cr \nonumber \\
&& \times \langle \psi_m(s) | \partial_s H(s) |\psi_n(s) \rangle.
\eea
If we choose $\Delta$ in the definition of $F$ to coincide with $\gamma$, then we see that 
\beq
i \partial_s |\psi_n(s)\rangle = J(s) |\psi_n(s)\rangle
\eeq
for all $n$, not just the ground state.  

Now, as we have repeatedly emphasized, there is no such gap $\gamma$ in a many-body localized phase.  However, there is typically an effective gap if we only consider states that can be connected to a given energy eigenstate by localized operators.  This expectation is confirmed, for example, in Ref. \cite{2007PhRvB..75o5111O}, which demonstrated numerically the absence of level repulsion. Hence, because $\partial_s H(s)$ is a sum of such localized operators, we can argue that although the filter function $F$ cannot remove all states nearby in energy (not even close), it can effectively remove states which have a sizable matrix element of $\partial_s H(s)$ with $|\psi_n\rangle$.  Of course, this may also fail in rare instances.  We leave it to future work to make further progress on the issue of stability.

\subsection{Application to the model}

Because our model is defined in terms of an arbitrary local unitary and a fixed point Hamiltonian, it immediately follows that there is a class of transformations under which the model is stable.  For example, we may change the random local unitary to another random local unitary and the model clearly remains localized.  Similarly, we may make small changes to energy level structure of the fixed point Hamiltonian and the phase will remain localized.  For example, when changing the local unitary from $U_1$ to $U_2$, the transformation is effected $U_2 U_1^{-1}$ which is still local.  It is interesting to ask how the quasi-adiabatic approach deals with such perturbations for which we know the exact answer.

Recall the definition of the model Hamiltonian $H$ in terms of $H_0$ and the unitary $U$:
\beq
H = U H_0 U^{-1}.
\eeq
Recall also that $U$ was the exponential of a local operator $K$:
\beq
U = e^{-i K}.
\eeq
Now introduce a family of Hamiltonians $H(s)$ defined by
\beq
H(s) = e^{- i s K} H_0 e^{i s K} = U(s) H_0 U^\dagger(s).
\eeq
We will test the machinery of quasi-adiabatic evolution on this family of Hamiltonians.

Since $H(s=0) = H_0$ we clearly begin in maximally localized phase where all energy eigenstates are product states.  We can calculate $\partial_s H(s)$ as
\beq
\partial_s H(s) = -i [K,H(s)] = -i U(s) [K,H_0] U^\dagger(s).
\eeq
Thus we see that the quasi-adiabatic generator is
\bea
&& U^\dagger(s) J(s) U(s) = \cr \nonumber \\
&& \left[\int dt F(t) e^{i H_0 t} [K,H_0] e^{-i H_0 t}\right].
\eea
To compute matrix elements of this operator with respect to the exact many-body eigenstates of $H(s)$, we use the fact that such eigenstates are related to eigenstates $|n\rangle$ of $H_0$ by $U(s)$.  Hence we have
\bea
&& (\langle n | U^\dagger(s)) J(s) (U(s) |m\rangle) = \cr \nonumber \\
&& \langle n | \int dt F(t) e^{i H_0 t} [K,H_0] e^{-i H_0 t} | m \rangle.
\eea
Using the properties of $F$ we see that for $|E(n)-E(m)| \geq \Delta$, the matrix elements are just those of $K$ itself.  Hence the quasi-adiabatic generator coincides for such states.  On the other hand, matrix elements of $J(s)$ do not reduce to those of $K$ for states with energies closer than $\Delta$.  

At this point, then, we see clearly the importance of localization physics.  The matrix elements are
\beq
\tilde{F}(E(n)-E(m)) (E(m)-E(n)) \langle n | K | m\rangle.
\eeq
$K$ is short-ranged and the $|n\rangle$ are all product states, hence the matrix elements $K(n,m) = \langle n |K |m\rangle$ can only be non-zero if $n$ and $m$ differ locally.  Yet the local spectrum of such a many-body localized phase is discrete, with an effective gap $\gamma$, hence almost all pairs of states with energy difference less than $\gamma$ will give a zero matrix element of $K$ (within a more general model, the matrix element might be exponentially small).  The final conclusion is that the quasi-adiabatic generator $J(s)$ is equivalent to $K$ for almost all states.  Since $K$ generates the exact unitary which transforms between eigenvectors for diffferent $s$, we see that within our model the quasi-adiabatic method leads to the correct conclusion that localization persists as we vary the Hamiltonian, e.g., each $H(s)$ is many-body localized.

\subsection{Level crossings}

We can also change the couplings entering the fixed point Hamiltonian without changing the operators which appear.  Consider $H_0(s)$ which is of the same form as $H_0$ but with variable couplings.  $H(s)$ is defined as $U H_0(s) U^{-1}$ with $U$ fixed.  It follows that $[\partial_s H(s) , H(s)]=0$ and hence $J(s)$ in this case is
\bea
&& -i J(s) = \int dt F(t) e^{i H(s) t} \partial_s H e^{-i H(s) t} \cr \nonumber 
&& = \left(\int dt F(t) \right) \partial_s H(s).
\eea
Since $\partial_s H$ has all the same eigenvectors as $H(s)$, it follows that $J(s)$ only generates an uninteresting phase for each energy eigenstate.  Hence in this case the resulting quasi-adiabatic transformation is of the form
\beq
\sum_n e^{i \theta(n,s)} U |n\rangle \langle n |U^\dagger 
\eeq
which manifestly preserves all eigenvectors up to a phase.

However, in a more generic physical system, one might expect avoided crossings over a very small energy window (see, for example, Ref. \cite{Altshuler13072010}).  For example, imagine we build up the Hamiltonian from the extreme localized using the family of Hamiltonians $H(s)$.  As $s$ is varied, there are bound to be very sharp avoided level crossings in the physical Hamiltonian.  This is because the matrix elements of the perturbation, while they are exponentially suppressed, may still connect states with an even smaller energy splitting in the limit of very large systems.  However, the quasi-adiabatic evolution will be fast on the scale of the induced mixing, so rather than follow the avoided crossing, the states will shoot through the crossing.  Given two initial states, of lower and higher energy, the quasi-adiabatic evolution will map low to high and vice versa instead of low to low as would occur if we ran the evolution for a long enough time.

However, this simple picture suggests that the quasi-adiabatic evolution will still map energy eigenstates to energy eigenstates, it may just get them ``out of order".  The locality of the qausi-adiabatic evolution combined with localized nature of all initial energy eigenstates means that all resulting evolved states are still localized.  If the evolved states are indeed close to energy eigenstates of the target Hamiltonian, then we have succeeded in demonstrating the stability of many-body localization.  Of course, what we have said is merely a sketch, since there are many closely spaced states and many-level crossings happening as the Hamiltonian is changed.  It remains to be seen what part of the above sketch can be made completely rigorous.

\section{Discussion}

In this paper we presented a simple model of a many-body localized phase.  The model describes a system deep within a many-body localized and is analogous to a fixed point of a real space renormalization scheme.  We showed that the model has a number of phenomenological features expected of many-body localized states.  All energy eigenstates are localized and obey an area law for entanglement entropy.  Energy is not transported even at finite energy density.  The growth of entanglement after a perturbation is also slow, either effectively not occurring at all or only growing logarithmically.  Finally, we gave some arguments supporting the hypothesis that our model is stable and hence describes a many-body localized phase with no relevant perturbations.

It has already been observed in one dimension that the low entanglement scaling of energy eigenstates and the slow growth of entanglement with time makes tensor network states an ideal tool to investigate the physics of many-body localized states.  Indeed, matrix product states and DMRG were used to great effect to study the entanglement dynamics of one dimensional localized phases \cite{2008PhRvB..77f4426Z,2012PhRvL.109a7202B}.  Our renormalization group circuit picture naturally fits into the context of more general tensor network states like the MERA \cite{mera}.  We have also proposed \cite{PhysRevD.86.065007,2012arXiv1209.3304S} that MERA provides a kind of holographic \cite{maldacena,polyakov,witten} description of ground states, so thinking along these lines, it would be interesting to find phases with features of a many-body localized state in the context of gravity.  Perhaps the disordered black holes of Ref. \cite{2011arXiv1102.2892A} could be useful.

There are also numerous open questions.  For example, it might be possible to rigorously prove or at least argue more completely that the model introduced above is stable.  The assumption of a single unitary transformation $U$ is clearly too extreme in the generic case (see, for example, the results of Ref. \cite{2013arXiv1306.5753B}), but it is not clear what assumption should replace it.  It should also be possible to give evidence for its stability using DMRG by checking the many-body spectrum for a variety of weak perturbations.  Another interesting question relates to the possibility of having some delocalized energy eigenstates coexisting with localized eigenstates.  Similarly, it would be interesting to try to understand a phase transition between a many-body localized phase and an ergodic phase where the system can serve as its own heat bath \cite{basko_loc,PhysRevB.82.174411}.  Finally, there is the question of whether the phenomena of many-body localization have any clear signature for experiments (beyond the absence of diffusion), for example, in cold atoms systems where a phonon heat bath can be excluded.

\textit{Acknowledgements:} We thank M.P.A. Fisher and A. Chandran for enlightening discussions which helped inspire the present model and J. Sau for earlier conversations about logarithmically growing entanglement arising from exponentially small interactions.  We also thank J. McGreevy, S. Gopalakrishnan, and M. Knap for discussions about many-body localization and Refs. \cite{2013arXiv1304.4605S,2013arXiv1305.4915H,2013arXiv1305.5554S}.  We thank A. Pal, D. Huse, and D. Abanin for helpful comments on the manuscript.  Significant progress on this work was made during ``Entanglement and Emergence II" at the Perimeter Institute, and we thank the Perimeter Institute for hospitality.  BGS is supported by a Simons Fellowship through Harvard University.

\bibliography{mbl_solvable}

\begin{thebibliography}{44}
\expandafter\ifx\csname natexlab\endcsname\relax\def\natexlab#1{#1}\fi
\expandafter\ifx\csname bibnamefont\endcsname\relax
  \def\bibnamefont#1{#1}\fi
\expandafter\ifx\csname bibfnamefont\endcsname\relax
  \def\bibfnamefont#1{#1}\fi
\expandafter\ifx\csname citenamefont\endcsname\relax
  \def\citenamefont#1{#1}\fi
\expandafter\ifx\csname url\endcsname\relax
  \def\url#1{\texttt{#1}}\fi
\expandafter\ifx\csname urlprefix\endcsname\relax\def\urlprefix{URL }\fi
\providecommand{\bibinfo}[2]{#2}
\providecommand{\eprint}[2][]{\url{#2}}

\bibitem[{\citenamefont{Anderson}(1958)}]{anderson_loc}
\bibinfo{author}{\bibfnamefont{P.~W.} \bibnamefont{Anderson}},
  \bibinfo{journal}{Phys. Rev.} \textbf{\bibinfo{volume}{109}},
  \bibinfo{pages}{1492} (\bibinfo{year}{1958}),
  \urlprefix\url{http://link.aps.org/doi/10.1103/PhysRev.109.1492}.

\bibitem[{\citenamefont{Lee and Ramakrishnan}(1985)}]{RevModPhys.57.287}
\bibinfo{author}{\bibfnamefont{P.~A.} \bibnamefont{Lee}} \bibnamefont{and}
  \bibinfo{author}{\bibfnamefont{T.~V.} \bibnamefont{Ramakrishnan}},
  \bibinfo{journal}{Rev. Mod. Phys.} \textbf{\bibinfo{volume}{57}},
  \bibinfo{pages}{287} (\bibinfo{year}{1985}),
  \urlprefix\url{http://link.aps.org/doi/10.1103/RevModPhys.57.287}.

\bibitem[{\citenamefont{Gornyi et~al.}(2005)\citenamefont{Gornyi, Mirlin, and
  Polyakov}}]{PhysRevLett.95.206603}
\bibinfo{author}{\bibfnamefont{I.~V.} \bibnamefont{Gornyi}},
  \bibinfo{author}{\bibfnamefont{A.~D.} \bibnamefont{Mirlin}},
  \bibnamefont{and} \bibinfo{author}{\bibfnamefont{D.~G.}
  \bibnamefont{Polyakov}}, \bibinfo{journal}{Phys. Rev. Lett.}
  \textbf{\bibinfo{volume}{95}}, \bibinfo{pages}{206603}
  (\bibinfo{year}{2005}),
  \urlprefix\url{http://link.aps.org/doi/10.1103/PhysRevLett.95.206603}.

\bibitem[{\citenamefont{{Basko} et~al.}(2006)\citenamefont{{Basko}, {Aleiner},
  and {Altshuler}}}]{basko_loc}
\bibinfo{author}{\bibfnamefont{D.~M.} \bibnamefont{{Basko}}},
  \bibinfo{author}{\bibfnamefont{I.~L.} \bibnamefont{{Aleiner}}},
  \bibnamefont{and} \bibinfo{author}{\bibfnamefont{B.~L.}
  \bibnamefont{{Altshuler}}}, \bibinfo{journal}{Annals of Physics}
  \textbf{\bibinfo{volume}{321}}, \bibinfo{pages}{1126} (\bibinfo{year}{2006}),
  \eprint{arXiv:cond-mat/0506617}.

\bibitem[{\citenamefont{{Oganesyan} and {Huse}}(2007)}]{2007PhRvB..75o5111O}
\bibinfo{author}{\bibfnamefont{V.}~\bibnamefont{{Oganesyan}}} \bibnamefont{and}
  \bibinfo{author}{\bibfnamefont{D.~A.} \bibnamefont{{Huse}}},
  \bibinfo{journal}{\prb} \textbf{\bibinfo{volume}{75}}, \bibinfo{eid}{155111}
  (\bibinfo{year}{2007}), \eprint{arXiv:cond-mat/0610854}.

\bibitem[{\citenamefont{{Oganesyan} et~al.}(2009)\citenamefont{{Oganesyan},
  {Pal}, and {Huse}}}]{2009PhRvB..80k5104O}
\bibinfo{author}{\bibfnamefont{V.}~\bibnamefont{{Oganesyan}}},
  \bibinfo{author}{\bibfnamefont{A.}~\bibnamefont{{Pal}}}, \bibnamefont{and}
  \bibinfo{author}{\bibfnamefont{D.~A.} \bibnamefont{{Huse}}},
  \bibinfo{journal}{\prb} \textbf{\bibinfo{volume}{80}}, \bibinfo{eid}{115104}
  (\bibinfo{year}{2009}), \eprint{0905.4112}.

\bibitem[{\citenamefont{Pal and Huse}(2010)}]{PhysRevB.82.174411}
\bibinfo{author}{\bibfnamefont{A.}~\bibnamefont{Pal}} \bibnamefont{and}
  \bibinfo{author}{\bibfnamefont{D.~A.} \bibnamefont{Huse}},
  \bibinfo{journal}{Phys. Rev. B} \textbf{\bibinfo{volume}{82}},
  \bibinfo{pages}{174411} (\bibinfo{year}{2010}),
  \urlprefix\url{http://link.aps.org/doi/10.1103/PhysRevB.82.174411}.

\bibitem[{\citenamefont{{{\v Z}nidari{\v c}} et~al.}(2008)\citenamefont{{{\v
  Z}nidari{\v c}}, {Prosen}, and {Prelov{\v s}ek}}}]{2008PhRvB..77f4426Z}
\bibinfo{author}{\bibfnamefont{M.}~\bibnamefont{{{\v Z}nidari{\v c}}}},
  \bibinfo{author}{\bibfnamefont{T.}~\bibnamefont{{Prosen}}}, \bibnamefont{and}
  \bibinfo{author}{\bibfnamefont{P.}~\bibnamefont{{Prelov{\v s}ek}}},
  \bibinfo{journal}{\prb} \textbf{\bibinfo{volume}{77}}, \bibinfo{eid}{064426}
  (\bibinfo{year}{2008}), \eprint{0706.2539}.

\bibitem[{\citenamefont{{Bardarson} et~al.}(2012)\citenamefont{{Bardarson},
  {Pollmann}, and {Moore}}}]{2012PhRvL.109a7202B}
\bibinfo{author}{\bibfnamefont{J.~H.} \bibnamefont{{Bardarson}}},
  \bibinfo{author}{\bibfnamefont{F.}~\bibnamefont{{Pollmann}}},
  \bibnamefont{and} \bibinfo{author}{\bibfnamefont{J.~E.}
  \bibnamefont{{Moore}}}, \bibinfo{journal}{Physical Review Letters}
  \textbf{\bibinfo{volume}{109}}, \bibinfo{eid}{017202} (\bibinfo{year}{2012}),
  \eprint{1202.5532}.

\bibitem[{\citenamefont{{Iyer} et~al.}(2013)\citenamefont{{Iyer}, {Oganesyan},
  {Refael}, and {Huse}}}]{2013PhRvB..87m4202I}
\bibinfo{author}{\bibfnamefont{S.}~\bibnamefont{{Iyer}}},
  \bibinfo{author}{\bibfnamefont{V.}~\bibnamefont{{Oganesyan}}},
  \bibinfo{author}{\bibfnamefont{G.}~\bibnamefont{{Refael}}}, \bibnamefont{and}
  \bibinfo{author}{\bibfnamefont{D.~A.} \bibnamefont{{Huse}}},
  \bibinfo{journal}{\prb} \textbf{\bibinfo{volume}{87}}, \bibinfo{eid}{134202}
  (\bibinfo{year}{2013}), \eprint{1212.4159}.

\bibitem[{\citenamefont{{D'Alessio} and
  {Polkovnikov}}(2013)}]{2013AnPhy.333...19D}
\bibinfo{author}{\bibfnamefont{L.}~\bibnamefont{{D'Alessio}}} \bibnamefont{and}
  \bibinfo{author}{\bibfnamefont{A.}~\bibnamefont{{Polkovnikov}}},
  \bibinfo{journal}{Annals of Physics} \textbf{\bibinfo{volume}{333}},
  \bibinfo{pages}{19} (\bibinfo{year}{2013}), \eprint{1210.2791}.

\bibitem[{\citenamefont{{Canovi} et~al.}(2012)\citenamefont{{Canovi},
  {Rossini}, {Fazio}, {Santoro}, and {Silva}}}]{2012NJPh...14i5020C}
\bibinfo{author}{\bibfnamefont{E.}~\bibnamefont{{Canovi}}},
  \bibinfo{author}{\bibfnamefont{D.}~\bibnamefont{{Rossini}}},
  \bibinfo{author}{\bibfnamefont{R.}~\bibnamefont{{Fazio}}},
  \bibinfo{author}{\bibfnamefont{G.~E.} \bibnamefont{{Santoro}}},
  \bibnamefont{and} \bibinfo{author}{\bibfnamefont{A.}~\bibnamefont{{Silva}}},
  \bibinfo{journal}{New Journal of Physics} \textbf{\bibinfo{volume}{14}},
  \bibinfo{eid}{095020} (\bibinfo{year}{2012}), \eprint{1205.0370}.

\bibitem[{\citenamefont{{De Luca} and
  {Scardicchio}}(2013)}]{2013EL....10137003D}
\bibinfo{author}{\bibfnamefont{A.}~\bibnamefont{{De Luca}}} \bibnamefont{and}
  \bibinfo{author}{\bibfnamefont{A.}~\bibnamefont{{Scardicchio}}},
  \bibinfo{journal}{EPL (Europhysics Letters)} \textbf{\bibinfo{volume}{101}},
  \bibinfo{pages}{37003} (\bibinfo{year}{2013}), \eprint{1206.2342}.

\bibitem[{\citenamefont{{Delande} et~al.}(2013)\citenamefont{{Delande},
  {Sacha}, {P{\l}odzie{\'n}}, {Avazbaev}, and
  {Zakrzewski}}}]{2013NJPh...15d5021D}
\bibinfo{author}{\bibfnamefont{D.}~\bibnamefont{{Delande}}},
  \bibinfo{author}{\bibfnamefont{K.}~\bibnamefont{{Sacha}}},
  \bibinfo{author}{\bibfnamefont{M.}~\bibnamefont{{P{\l}odzie{\'n}}}},
  \bibinfo{author}{\bibfnamefont{S.~K.} \bibnamefont{{Avazbaev}}},
  \bibnamefont{and}
  \bibinfo{author}{\bibfnamefont{J.}~\bibnamefont{{Zakrzewski}}},
  \bibinfo{journal}{New Journal of Physics} \textbf{\bibinfo{volume}{15}},
  \bibinfo{eid}{045021} (\bibinfo{year}{2013}), \eprint{1207.2001}.

\bibitem[{\citenamefont{{Huse} et~al.}(2013)\citenamefont{{Huse},
  {Nandkishore}, {Oganesyan}, {Pal}, and {Sondhi}}}]{2013arXiv1304.1158H}
\bibinfo{author}{\bibfnamefont{D.~A.} \bibnamefont{{Huse}}},
  \bibinfo{author}{\bibfnamefont{R.}~\bibnamefont{{Nandkishore}}},
  \bibinfo{author}{\bibfnamefont{V.}~\bibnamefont{{Oganesyan}}},
  \bibinfo{author}{\bibfnamefont{A.}~\bibnamefont{{Pal}}}, \bibnamefont{and}
  \bibinfo{author}{\bibfnamefont{S.~L.} \bibnamefont{{Sondhi}}},
  \bibinfo{journal}{ArXiv e-prints}  (\bibinfo{year}{2013}),
  \eprint{1304.1158}.

\bibitem[{\citenamefont{Serbyn et~al.}(2013)\citenamefont{Serbyn,
  Papi\ifmmode~\acute{c}\else \'{c}\fi{}, and Abanin}}]{2013arXiv1304.4605S}
\bibinfo{author}{\bibfnamefont{M.}~\bibnamefont{Serbyn}},
  \bibinfo{author}{\bibfnamefont{Z.}~\bibnamefont{Papi\ifmmode~\acute{c}\else
  \'{c}\fi{}}}, \bibnamefont{and} \bibinfo{author}{\bibfnamefont{D.~A.}
  \bibnamefont{Abanin}}, \bibinfo{journal}{Phys. Rev. Lett.}
  \textbf{\bibinfo{volume}{110}}, \bibinfo{pages}{260601}
  (\bibinfo{year}{2013}),
  \urlprefix\url{http://link.aps.org/doi/10.1103/PhysRevLett.110.260601}.

\bibitem[{\citenamefont{{Huse} and {Oganesyan}}(2013)}]{2013arXiv1305.4915H}
\bibinfo{author}{\bibfnamefont{D.~A.} \bibnamefont{{Huse}}} \bibnamefont{and}
  \bibinfo{author}{\bibfnamefont{V.}~\bibnamefont{{Oganesyan}}},
  \bibinfo{journal}{ArXiv e-prints}  (\bibinfo{year}{2013}),
  \eprint{1305.4915}.

\bibitem[{\citenamefont{{Serbyn} et~al.}(2013)\citenamefont{{Serbyn},
  {Papi{\'c}}, and {Abanin}}}]{2013arXiv1305.5554S}
\bibinfo{author}{\bibfnamefont{M.}~\bibnamefont{{Serbyn}}},
  \bibinfo{author}{\bibfnamefont{Z.}~\bibnamefont{{Papi{\'c}}}},
  \bibnamefont{and} \bibinfo{author}{\bibfnamefont{D.~A.}
  \bibnamefont{{Abanin}}}, \bibinfo{journal}{ArXiv e-prints}
  (\bibinfo{year}{2013}), \eprint{1305.5554}.

\bibitem[{\citenamefont{{Bauer} and {Nayak}}(2013)}]{2013arXiv1306.5753B}
\bibinfo{author}{\bibfnamefont{B.}~\bibnamefont{{Bauer}}} \bibnamefont{and}
  \bibinfo{author}{\bibfnamefont{C.}~\bibnamefont{{Nayak}}},
  \bibinfo{journal}{ArXiv e-prints}  (\bibinfo{year}{2013}),
  \eprint{1306.5753}.

\bibitem[{\citenamefont{Vidal}(2008)}]{mera}
\bibinfo{author}{\bibfnamefont{G.}~\bibnamefont{Vidal}},
  \bibinfo{journal}{Phys. Rev. Lett.} \textbf{\bibinfo{volume}{101}},
  \bibinfo{pages}{110501} (\bibinfo{year}{2008}).

\bibitem[{\citenamefont{Verstraete et~al.}(2008)\citenamefont{Verstraete,
  Cirac, and Murg}}]{peps}
\bibinfo{author}{\bibfnamefont{F.}~\bibnamefont{Verstraete}},
  \bibinfo{author}{\bibfnamefont{J.}~\bibnamefont{Cirac}}, \bibnamefont{and}
  \bibinfo{author}{\bibfnamefont{V.}~\bibnamefont{Murg}},
  \bibinfo{journal}{Adv. Phys.} \textbf{\bibinfo{volume}{57}},
  \bibinfo{pages}{143} (\bibinfo{year}{2008}).

\bibitem[{\citenamefont{Gu et~al.}(2008)\citenamefont{Gu, Levin, and
  Wen}}]{terg}
\bibinfo{author}{\bibfnamefont{Z.-C.} \bibnamefont{Gu}},
  \bibinfo{author}{\bibfnamefont{M.}~\bibnamefont{Levin}}, \bibnamefont{and}
  \bibinfo{author}{\bibfnamefont{X.-G.} \bibnamefont{Wen}},
  \bibinfo{journal}{Phys. Rev. B} \textbf{\bibinfo{volume}{78}},
  \bibinfo{pages}{205116} (\bibinfo{year}{2008}).

\bibitem[{\citenamefont{White}(1992)}]{PhysRevLett.69.2863}
\bibinfo{author}{\bibfnamefont{S.~R.} \bibnamefont{White}},
  \bibinfo{journal}{Phys. Rev. Lett.} \textbf{\bibinfo{volume}{69}},
  \bibinfo{pages}{2863} (\bibinfo{year}{1992}),
  \urlprefix\url{http://link.aps.org/doi/10.1103/PhysRevLett.69.2863}.

\bibitem[{\citenamefont{Schollw\"ock}(2005)}]{dmrg_review}
\bibinfo{author}{\bibfnamefont{U.}~\bibnamefont{Schollw\"ock}},
  \bibinfo{journal}{Rev. Mod. Phys.} \textbf{\bibinfo{volume}{77}},
  \bibinfo{pages}{259} (\bibinfo{year}{2005}).

\bibitem[{\citenamefont{Vidal}(2007)}]{vidal_er}
\bibinfo{author}{\bibfnamefont{G.}~\bibnamefont{Vidal}},
  \bibinfo{journal}{Phys. Rev. Lett.} \textbf{\bibinfo{volume}{99}},
  \bibinfo{pages}{220405} (\bibinfo{year}{2007}).

\bibitem[{\citenamefont{{Montangero} et~al.}(2009)\citenamefont{{Montangero},
  {Rizzi}, {Giovannetti}, and {Fazio}}}]{2009PhRvB..80k3103M}
\bibinfo{author}{\bibfnamefont{S.}~\bibnamefont{{Montangero}}},
  \bibinfo{author}{\bibfnamefont{M.}~\bibnamefont{{Rizzi}}},
  \bibinfo{author}{\bibfnamefont{V.}~\bibnamefont{{Giovannetti}}},
  \bibnamefont{and} \bibinfo{author}{\bibfnamefont{R.}~\bibnamefont{{Fazio}}},
  \bibinfo{journal}{\prb} \textbf{\bibinfo{volume}{80}}, \bibinfo{eid}{113103}
  (\bibinfo{year}{2009}), \eprint{0810.1414}.

\bibitem[{\citenamefont{{Pfeifer} et~al.}(2009)\citenamefont{{Pfeifer},
  {Evenbly}, and {Vidal}}}]{2009PhRvA..79d0301P}
\bibinfo{author}{\bibfnamefont{R.~N.~C.} \bibnamefont{{Pfeifer}}},
  \bibinfo{author}{\bibfnamefont{G.}~\bibnamefont{{Evenbly}}},
  \bibnamefont{and} \bibinfo{author}{\bibfnamefont{G.}~\bibnamefont{{Vidal}}},
  \bibinfo{journal}{\pra} \textbf{\bibinfo{volume}{79}}, \bibinfo{eid}{040301}
  (\bibinfo{year}{2009}), \eprint{0810.0580}.

\bibitem[{\citenamefont{Gu et~al.}(2009)\citenamefont{Gu, Levin, Swingle, and
  Wen}}]{PhysRevB.79.085118}
\bibinfo{author}{\bibfnamefont{Z.-C.} \bibnamefont{Gu}},
  \bibinfo{author}{\bibfnamefont{M.}~\bibnamefont{Levin}},
  \bibinfo{author}{\bibfnamefont{B.}~\bibnamefont{Swingle}}, \bibnamefont{and}
  \bibinfo{author}{\bibfnamefont{X.-G.} \bibnamefont{Wen}},
  \bibinfo{journal}{Phys. Rev. B} \textbf{\bibinfo{volume}{79}},
  \bibinfo{pages}{085118} (\bibinfo{year}{2009}),
  \urlprefix\url{http://link.aps.org/doi/10.1103/PhysRevB.79.085118}.

\bibitem[{\citenamefont{Buerschaper et~al.}(2009)\citenamefont{Buerschaper,
  Aguado, and Vidal}}]{PhysRevB.79.085119}
\bibinfo{author}{\bibfnamefont{O.}~\bibnamefont{Buerschaper}},
  \bibinfo{author}{\bibfnamefont{M.}~\bibnamefont{Aguado}}, \bibnamefont{and}
  \bibinfo{author}{\bibfnamefont{G.}~\bibnamefont{Vidal}},
  \bibinfo{journal}{Phys. Rev. B} \textbf{\bibinfo{volume}{79}},
  \bibinfo{pages}{085119} (\bibinfo{year}{2009}),
  \urlprefix\url{http://link.aps.org/doi/10.1103/PhysRevB.79.085119}.

\bibitem[{\citenamefont{Swingle}(2012)}]{PhysRevD.86.065007}
\bibinfo{author}{\bibfnamefont{B.}~\bibnamefont{Swingle}},
  \bibinfo{journal}{Phys. Rev. D} \textbf{\bibinfo{volume}{86}},
  \bibinfo{pages}{065007} (\bibinfo{year}{2012}),
  \urlprefix\url{http://link.aps.org/doi/10.1103/PhysRevD.86.065007}.

\bibitem[{\citenamefont{{Swingle}}(2012)}]{2012arXiv1209.3304S}
\bibinfo{author}{\bibfnamefont{B.}~\bibnamefont{{Swingle}}},
  \bibinfo{journal}{ArXiv e-prints}  (\bibinfo{year}{2012}),
  \eprint{1209.3304}.

\bibitem[{\citenamefont{Maldacena}(1998)}]{maldacena}
\bibinfo{author}{\bibfnamefont{J.~M.} \bibnamefont{Maldacena}},
  \bibinfo{journal}{Adv. Theor. Math. Phys.} \textbf{\bibinfo{volume}{2}},
  \bibinfo{pages}{231} (\bibinfo{year}{1998}).

\bibitem[{\citenamefont{Gubser et~al.}(1998)\citenamefont{Gubser, Klebanov, and
  Polyakov}}]{polyakov}
\bibinfo{author}{\bibfnamefont{S.~S.} \bibnamefont{Gubser}},
  \bibinfo{author}{\bibfnamefont{I.~R.} \bibnamefont{Klebanov}},
  \bibnamefont{and} \bibinfo{author}{\bibfnamefont{A.~M.}
  \bibnamefont{Polyakov}}, \bibinfo{journal}{Phys. Lett. B}
  \textbf{\bibinfo{volume}{428}}, \bibinfo{pages}{105} (\bibinfo{year}{1998}).

\bibitem[{\citenamefont{Witten}(1998)}]{witten}
\bibinfo{author}{\bibfnamefont{E.}~\bibnamefont{Witten}},
  \bibinfo{journal}{Adv. Theor. Math. Phys.} \textbf{\bibinfo{volume}{2}},
  \bibinfo{pages}{253} (\bibinfo{year}{1998}).

\bibitem[{\citenamefont{{Monthus} and {Garel}}(2010)}]{2010PhRvB..81m4202M}
\bibinfo{author}{\bibfnamefont{C.}~\bibnamefont{{Monthus}}} \bibnamefont{and}
  \bibinfo{author}{\bibfnamefont{T.}~\bibnamefont{{Garel}}},
  \bibinfo{journal}{\prb} \textbf{\bibinfo{volume}{81}}, \bibinfo{eid}{134202}
  (\bibinfo{year}{2010}), \eprint{1001.2984}.

\bibitem[{\citenamefont{{Vosk} and {Altman}}(2013)}]{2013PhRvL.110f7204V}
\bibinfo{author}{\bibfnamefont{R.}~\bibnamefont{{Vosk}}} \bibnamefont{and}
  \bibinfo{author}{\bibfnamefont{E.}~\bibnamefont{{Altman}}},
  \bibinfo{journal}{Physical Review Letters} \textbf{\bibinfo{volume}{110}},
  \bibinfo{eid}{067204} (\bibinfo{year}{2013}), \eprint{1205.0026}.

\bibitem[{\citenamefont{Lieb and Robinson}(1972)}]{LRbound}
\bibinfo{author}{\bibfnamefont{E.}~\bibnamefont{Lieb}} \bibnamefont{and}
  \bibinfo{author}{\bibfnamefont{D.}~\bibnamefont{Robinson}},
  \bibinfo{journal}{Communications in Mathematical Physics}
  \textbf{\bibinfo{volume}{28}}, \bibinfo{pages}{251} (\bibinfo{year}{1972}),
  ISSN \bibinfo{issn}{0010-3616},
  \urlprefix\url{http://dx.doi.org/10.1007/BF01645779}.

\bibitem[{\citenamefont{D\"ur et~al.}(2001)\citenamefont{D\"ur, Vidal, Cirac,
  Linden, and Popescu}}]{ent_cap_HAHB}
\bibinfo{author}{\bibfnamefont{W.}~\bibnamefont{D\"ur}},
  \bibinfo{author}{\bibfnamefont{G.}~\bibnamefont{Vidal}},
  \bibinfo{author}{\bibfnamefont{J.~I.} \bibnamefont{Cirac}},
  \bibinfo{author}{\bibfnamefont{N.}~\bibnamefont{Linden}}, \bibnamefont{and}
  \bibinfo{author}{\bibfnamefont{S.}~\bibnamefont{Popescu}},
  \bibinfo{journal}{Phys. Rev. Lett.} \textbf{\bibinfo{volume}{87}},
  \bibinfo{pages}{137901} (\bibinfo{year}{2001}),
  \urlprefix\url{http://link.aps.org/doi/10.1103/PhysRevLett.87.137901}.

\bibitem[{\citenamefont{Eisert and Osborne}(2006)}]{PhysRevLett.97.150404}
\bibinfo{author}{\bibfnamefont{J.}~\bibnamefont{Eisert}} \bibnamefont{and}
  \bibinfo{author}{\bibfnamefont{T.~J.} \bibnamefont{Osborne}},
  \bibinfo{journal}{Phys. Rev. Lett.} \textbf{\bibinfo{volume}{97}},
  \bibinfo{pages}{150404} (\bibinfo{year}{2006}),
  \urlprefix\url{http://link.aps.org/doi/10.1103/PhysRevLett.97.150404}.

\bibitem[{\citenamefont{Bravyi et~al.}(2006)\citenamefont{Bravyi, Hastings, and
  Verstraete}}]{PhysRevLett.97.050401}
\bibinfo{author}{\bibfnamefont{S.}~\bibnamefont{Bravyi}},
  \bibinfo{author}{\bibfnamefont{M.~B.} \bibnamefont{Hastings}},
  \bibnamefont{and}
  \bibinfo{author}{\bibfnamefont{F.}~\bibnamefont{Verstraete}},
  \bibinfo{journal}{Phys. Rev. Lett.} \textbf{\bibinfo{volume}{97}},
  \bibinfo{pages}{050401} (\bibinfo{year}{2006}),
  \urlprefix\url{http://link.aps.org/doi/10.1103/PhysRevLett.97.050401}.

\bibitem[{\citenamefont{Altshuler et~al.}(2010)\citenamefont{Altshuler, Krovi,
  and Roland}}]{Altshuler13072010}
\bibinfo{author}{\bibfnamefont{B.}~\bibnamefont{Altshuler}},
  \bibinfo{author}{\bibfnamefont{H.}~\bibnamefont{Krovi}}, \bibnamefont{and}
  \bibinfo{author}{\bibfnamefont{J.}~\bibnamefont{Roland}},
  \bibinfo{journal}{Proceedings of the National Academy of Sciences}
  \textbf{\bibinfo{volume}{107}}, \bibinfo{pages}{12446}
  (\bibinfo{year}{2010}),
  \eprint{http://www.pnas.org/content/107/28/12446.full.pdf+html},
  \urlprefix\url{http://www.pnas.org/content/107/28/12446.abstract}.

\bibitem[{\citenamefont{Hastings and Wen}(2005)}]{quasiadiabatic}
\bibinfo{author}{\bibfnamefont{M.~B.} \bibnamefont{Hastings}} \bibnamefont{and}
  \bibinfo{author}{\bibfnamefont{X.-G.} \bibnamefont{Wen}},
  \bibinfo{journal}{Phys. Rev. B} \textbf{\bibinfo{volume}{72}},
  \bibinfo{pages}{045141} (\bibinfo{year}{2005}),
  \urlprefix\url{http://link.aps.org/doi/10.1103/PhysRevB.72.045141}.

\bibitem[{\citenamefont{{Hastings}}(2010)}]{2010arXiv1001.5280H}
\bibinfo{author}{\bibfnamefont{M.~B.} \bibnamefont{{Hastings}}},
  \bibinfo{journal}{ArXiv e-prints}  (\bibinfo{year}{2010}),
  \eprint{1001.5280}.

\bibitem[{\citenamefont{{Adams} and {Yaida}}(2011)}]{2011arXiv1102.2892A}
\bibinfo{author}{\bibfnamefont{A.}~\bibnamefont{{Adams}}} \bibnamefont{and}
  \bibinfo{author}{\bibfnamefont{S.}~\bibnamefont{{Yaida}}},
  \bibinfo{journal}{ArXiv e-prints}  (\bibinfo{year}{2011}),
  \eprint{1102.2892}.

\end{thebibliography}

\end{document}